# Measurements of Gated Conjugated Polymer with Electrode Spacing Down to Several Nanometers

David Abusch-Magder[†], Takao Someya[§], James Wang, Edward Laskowski, Ananth Dodabalapur[†], Zhenan Bao, D.M. Tennant

Bell Labs, Lucent Technologies, 600 Mountain Ave, Murray Hill NJ 07974

Abstract:

In this letter we describe electronic measurements of a conjugated polymer of phenylenevinylene (PPV) with electrode spacings down to 20 nm; all measurements are made in a gated transistor geometry. With rectangular electrodes we find that the current is fit by an exponential in the applied electric field for spacings between 50 nm and 2μm. Based on this finding we conclude that the current is not injection limited, and is concentrated in a very small region; we also discuss possible transport mechanisms. The calculated mobility appears exponential in the electric field rather than in the square root of field. We also show fabricated triangular electrodes with spacings down to 5 nm, and discuss measurements with spacings down to 20 nm in which a single chain of polymer may dominate the conductance.

---

[†] email: davidam@bell-labs.com
[§] Also at Department of Electrical Engineering, Columbia University, New York, NY 10027
[†] Currently at Department of Electrical and Computer Engineering, University of Texas, Austin, TX 78712

Conjugated polymers have been studied in depth since their discovery more than twenty years ago.[1]  Much attention has been paid to their optical and conducting properties because of their scientific interest and their possible use in organic light emitting diodes[2] as well as novel electronic applications.[3]  Conjugated polymers have been discussed, along with other one dimensional conductors, as potential wires for molecular electronic devices.[4]  While understanding of their electronic properties at the nanometer scale is clearly crucial, previous reports of organic transistors with submicron electrode spacing[5,6] have not focused on conjugated polymers.

In this letter we describe the fabrication and measurement of two sets of gated devices with electrode spacing between 5 nm and 2 μm which can be used to study transport in polymeric and organic films; we use these devices to study hole transport in a poly(phenylene vinylene) (PPV) derivative.  While PPV has been extensively studied due to its light emitting properties,[2] previous work on PPV has been in two terminal diode structures[7] with electrode spacings >100 nm.  Here we report that in devices with rectangular electrodes the current and consequently mobility is exponentially dependent on the applied electric field at high fields.  Measurements of triangular electrode structures suggest that in some devices a single chain of polymer dominates the conductance.

The two different device geometries we have used to investigate the electronic properties of conjugated polymers are shown in figures 1a and 3a&b.  All devices were fabricated on a heavily doped $n^+$ silicon wafer ($\rho$=2 – 6 mΩ-cm) with 950 Å of thermal $SiO_2$ grown to insulate the substrate. The electrodes were patterned by electron-beam lithography using a 500 Å film of 950 kDa molecular weight poly(methylmethacrylate), PMMA, as a resist. The PMMA was spin cast on the wafer, exposed using a JEOL JBX 9300FS e-beam lithography system operating at 100 kV at doses of 1050 $\mu$C/cm$^2$, and developed in a solution of methyl *iso*-butyl ketone: *iso*propanol (1:2). A thin layer of metal consisting of 20 Å of Ti and 50 Å of $Au_{0.6}Pd_{0.4}$ was evaporated and then lifted off in acetone. The resulting electrode structures appeared very smooth under examination by scanning electron microscopy (SEM) as can be seen in figure 1a and had a sheet resistance of $\rho \approx 90$ $\Omega/\square$. Contact to the e-beam patterned electrode structures was made by forming overlapping pads of ~(50 $\mu$m)$^2$ using conventional contact photolithography and liftoff. These large pads were readily contacted with a probe station.

The conjugated polymer used in our study, which we will call PPV-7, is a solubilized derivative of poly(phenylene vinylene), PPV, with heptyloxy side chains;[8] its chemical structure is shown in figure 1b. The polydispersity of the PPV-7 is 2.7, and the number average molecular weight is 18.5 kDa, determined by Gel Permeation Chromatography using monodispersed polystyrenes as standards, corresponding to a number average length of 55 nm; the weight average molecular weight is 50 kDa yielding an average chain length of 150 nm. For this study a fresh solution of 1.9 mg PPV-7 / ml of toluene was prepared and deposited on the electrode structure using drop casting. We placed the sample in a petri dish covered by a watch glass in a nitrogen environment thus extending the drying process over several days in an ambient of toluene and dry nitrogen. We measured the electrical characteristics of the dried samples in a probe station at 313 K under flowing dry nitrogen with minimal ambient illumination using an HP4155 with a current noise floor of ≈100 fA.

We begin our discussion of the experimental results by examining the current–voltage characteristics of a set of rectangular devices which have a channel length ℓ much shorter than the device width $w$; a micrograph of one of these rectangular devices is shown in figure 1a. The current through the transistor is linearly dependent on gate voltage, with a threshold for hole conduction of ≈15 V (see inset figure 1c). Above threshold the current $I$ is fit over three orders-of-magnitude by $I = C \times \exp\left[(V - V_1)/V_0\right]$ as seen in figure 1c; here $C$=1pA, while $V_0$ and $V_1$ are fitting parameters. All of the devices of this type are fit well using this equation. The strong dependence of current on $V_{sd}$ is uncharacteristic of conventional field effect transistors but is consistent with previous measurements made on PPV materials;[7] this unconventional behavior makes it difficult to extract standard transistor characteristics.

The fitting parameters for all our rectangular devices of different channel lengths are shown in figure 2a (left axis). The most striking observation is that the fitting parameter $V_0$ is linearly dependent on ℓ from 50 nm up to 2 μm, indicating that the current is exponentially dependent on the electric field $E \equiv V_{sd}/\ell$. Thus $I = C \times \exp\left[(E - E_1)/E_0\right]$, where $C$=1 pA and $E_0 = V_0/\ell \approx 16\text{mV/nm}$, and $E_1$ is plotted in figure 2a (right axis). This result yields significant insights into both the pattern of current flow and the underlying conduction mechanism. The exponential dependence of current on electric field implies that the current in these devices is tightly concentrated at the points of highest field. Since in the plane of the electrodes $E_{max} \gg E_0$, the current falls exponentially above that plane as the field weakens. Furthermore, the dependence of current on electric field between the electrodes over nearly two decades in channel length argues strongly that the current is limited by transport through the material between the electrodes; this precludes injection limited current. The linear modulation of current down to $I$=0 by the gate voltage (inset figure 1c) demonstrates that the current is due to gate modulated charge carriers that flow in the plane of the electrodes. The electrostatics of this devices more closely resemble a transistor than a space charge limited device[7,9] despite the novel current-voltage characteristics. These conclusions are consistent with the device geometry as well as the exponential decay of current above the electrodes. Finally we focus on the dependence of $E_1$ on $\ell$. Figure 2a shows that in the longer devices ($\ell$>300 nm) the value of $E_1$ is approximately constant; at shorter lengths $E_1$ increases, indicating a larger field is necessary to achieve 1 pA of current. This length scale may be related to the chain length of the polymer or to a crossover in the electrostatics as $\ell$ becomes longer than the $SiO_2$ thickness.

The finding that current varies exponentially with electric field also has important implications to the transport mechanism of the PPV-7. The exponential field dependence can not be explained by tunneling or thermal activation across a single barrier over such a wide range of channel lengths. Similar behavior has been observed in amorphous semiconductors[10] and in nanometer scale systems consisting of arrays of semiconducting nanoparticles.[11] In all these cases it is believed that the current path consists of many barriers in series (either hopping or tunneling). During current flow the total voltage across the device is divided among these barriers. If the current through an individual barrier is exponentially dependent on the voltage across it, a series of such barriers results in a current exponentially dependent on field. This dependence on field will be independent of the length scale as long as the device is much longer than the mean barrier separation.[9] In our study the exact nature of the barriers to conduction is not known, however two possibilities seem likely. First, the current may be limited by transport from one PPV-7 chain to its neighbor; this interchain transport is presumably limited by tunneling across the heptyloxy sidegroups. Alternatively, disorder along the chain may cause on-chain charge localization;[12] the electron-electron interactions between these small conducting regions might then enhance the barrier to charge transport (Coulomb blockade) as it does in arrays of nanoparticles.[11]

Our results can be interpreted as a variation of the mobility of PPV-7 with electric field. There is a long history of strong field dependence of the mobility in PPV materials[7,13] as well as in other organic conductors.[14] Our results would then predict that the mobility, $\mu$, is given by[15] $\ln \mu = E/E_0$ as shown in figure 2b. Using microwave measurements, Martens $et\ al.$[9] found a similar functional dependence of PPV mobility on field at lengths of 5 nm; however, at longer length scales they found $\ln \mu \propto \sqrt{E}$. Our data are not well described by the latter form. The discrepancy between the results may be explained if our device probes the conductance along the length of a polymer chain while their measurements are dominated by interchain transport. Further nanometer scale measurements in the transistor geometry will be necessary to resolve this issue.

Finally we present data for a different class of devices consisting of two triangular electrodes with separations down to 20 nm (see micrograph in figure 3a). These triangular devices also show current that is exponential in the applied voltage (see figure 3c). However their $I(V_{sd})$ behavior is more complicated, crossing over from linear $I(V_{sd})$ at low voltage to the familiar exponential curve at higher voltages. Because the geometry of the triangular devices is less symmetric it is difficult to estimate the fields present, and therefore to make a direct comparison to the rectangular devices; we plan to use modeling to facilitate the comparison.[16] Despite the challenges of determining the fields in this device it is clear that both the thinness of the electrodes and the small radius of curvature at their vertices concentrate the field. Given our findings above we conclude that the current flows dominantly in a very small area of the device between the vertices of the triangles. Remarkably some of the triangular devices show much larger currents than the rectangular devices of the same spacing,[17] while others show no measurable current even with an applied voltage of 7 V. Several of the triangular devices have switched during the course of measurement to a state of high current before switching back to a lower current state. Such switching is characteristic of mesoscopic systems where the current flows through one or several paths and is therefore strongly affected by changes along those paths.[18] Thus switching in this system suggests that the current is flowing through a few, or possibly even one chain of PPV-7. Further experiments should help clarify this issue.

In conclusion we have fabricated gated transistor structures with separations down to 5 nm for studying the nanometer-scale conducting properties of polymers, and have used these devices to study the polymer PPV-7. Rectangular devices, with spacings from 50 nm up to 2 μm, show current and consequently mobility that are exponentially dependent on the applied electric field. We have concluded that current is strongly concentrated in these devices because of the current-field relationship, and suggested some possible microscopic mechanisms that might account for such behavior. Finally we presented results from triangular electrode structures in some of which a single chain of polymer may dominate the conductance.

**Acknowledgments:**

We gratefully acknowledge discussions with Don Monroe, Steve Simon, Sharad Ramanathan, M. Ashraf Alam, David Goldhaber-Gordon, Bert de Boer and Nikolai Zhitenev. TS acknowledges financial support from JSPS.

**Figure Captions:**

**Figure 1:** (a) Shows a micrograph of one of the devices with rectangular electrodes. The channel region has $\ell$=50 nm, and $w$=1 μm. (b) Shows the chemical structure of the conjugated polymer PPV-7 used in this study. The repeating unit is approximately 1.3 nm long. (c) The $I(V_{sd})$ curve for a rectangular device with $\ell$=150 nm, $w$=5 μm, and $V_g$=-20 V is shown along with a fit to an exponential curve. The inset shows the same device's gate voltage dependence at $V_{sd}$=-10 V from which we conclude that the threshold is ≈15 V.

**Figure 2:** (a) The left axis shows the exponential fitting parameter $V_0$ used in fitting the $I(V_{sd})$ curve as a function of channel length $\ell$. The linear dependence indicates that the electric field between the electrodes determines the current. The solid line is a fit to the data with a slope of $E_0$≈16 mV/nm. The right axis plots the fitting parameter $E_1$ (plotted with ×) as a function of $\ell$; the dashed curve, which becomes constant for $\ell$>300 nm, is a guide to the eye. Because $E_0$ shows no significant gate voltage dependence, the data at gate voltages between -10 V and -40 V are shown together. (b) Shows the mobility, μ, as a function of electric field, $E$, for the device shown in figure 1c. As discussed in the text, these data are not consistent with $\ln \mu \propto \sqrt{E}$ as reported in the literature, but are fit well by $\ln \mu = E / E_0$.

**Figure 3:** (a) A SEM of one such device with an electrode spacing of ≈25 nm and tip radius of curvature of 10 nm. (b) A second SEM shows a similar device with 50 Å thick

Au$_{0.6}$Pd$_{0.4}$ electrodes which has a gap of ≈5 nm; this micrograph demonstrates the potential for this fabrication scheme to produce ultrashort devices, though no measurements of PPV-7 were conducted with this device. (c) Shows the $I(V_{sd})$ curve of a device with triangular electrodes. At higher $V_{sd}$ the current is exponential in applied voltage, but at $|V_{sd}| < 0.25$V the current is linear with $R$=23 GΩ. The data were taken at $V_g$=-20 V on a device which has a gap of ≈35 nm and a tip radius of curvature of ≈10 nm.

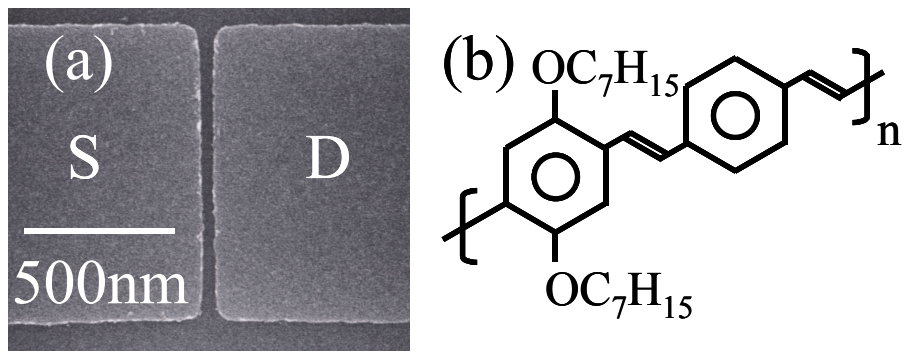

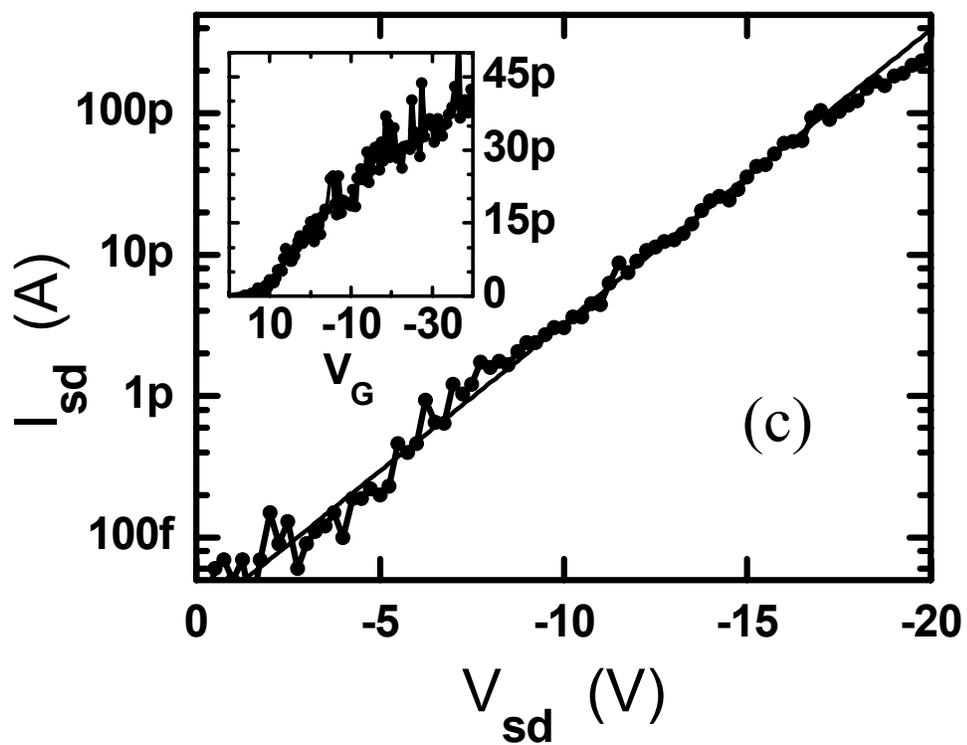

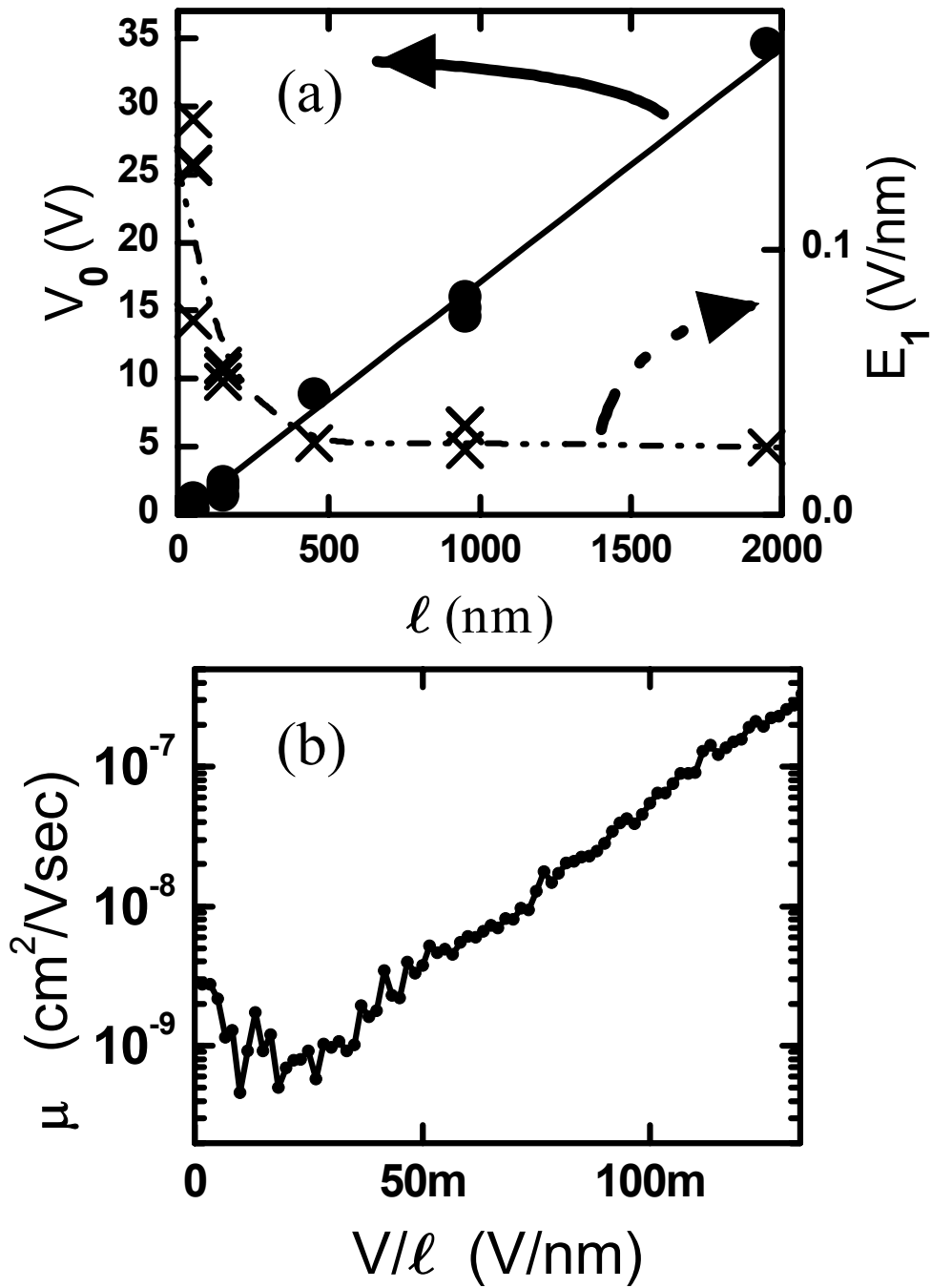

Abusch-Magder, *et al.*, APL, figure 2

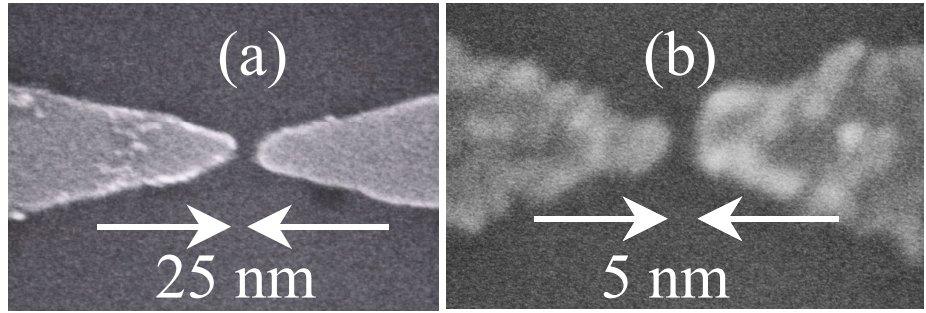

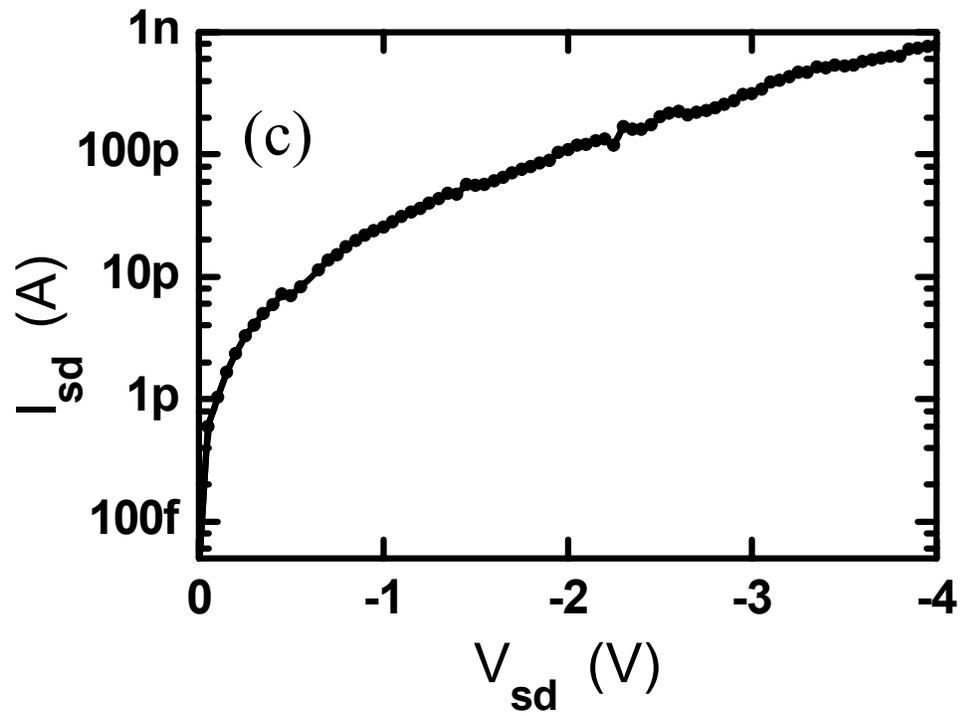

Abusch-Magder, *et al.*, APL, figure 3